\documentclass[aps,pra,twocolumn,nofootinbib]{revtex4-1}

\usepackage{graphicx,amsmath,amssymb, color}

\begin{document}

\title{Experimental demonstration of optimal probabilistic  enhancement \\ of quantum coherence }

\author{Robert St\'{a}rek}
\affiliation{Department of Optics, Palack\' y University, 17. listopadu 1192/12,  771~46 Olomouc,  Czech Republic}

\author{Michal Mi\v{c}uda}
\affiliation{Department of Optics, Palack\' y University, 17. listopadu 1192/12,  771~46 Olomouc,  Czech Republic}

\author{Michal Kol\'{a}\v{r}}
\affiliation{Department of Optics, Palack\' y University, 17. listopadu 1192/12,  771~46 Olomouc,  Czech Republic}

\author{Radim Filip}
\affiliation{Department of Optics, Palack\' y University, 17. listopadu 1192/12,  771~46 Olomouc,  Czech Republic}

\author{Jarom\' ir Fiur\' a\v sek}
\affiliation{Department of Optics, Palack\' y University, 17. listopadu 1192/12,  771~46 Olomouc,  Czech Republic}

\begin{abstract}
We theoretically and experimentally investigate conditional enhancement of overall coherence of quantum states by probabilistic quantum operations that apply to the input state a quantum filter diagonal in the basis of incoherent states. 
We identify the optimal filters that for a given probability of successful filtering maximize the output coherence. 
We verify the performance of the studied quantum filters in a proof-of-principle experiment with linear optics, where a pair of two-level quantum systems is represented by polarization states of two photons. 
We comprehensively characterize the implemented two-qubit linear optical quantum filters by full quantum process tomography and we experimentally observe the optimal quantum coherence enhancement by quantum filtering. 

\end{abstract}

\maketitle

\section{Introduction}

Quantum coherence is a valuable resource \cite{Baumgratz2014,Winter2016,Streltsov2017,Chitambar2019} in many areas of quantum science and technology, such as quantum information processing or quantum metrology \cite{Giovanetti2011,Toth2014}.
Of particular interest is the role of quantum coherence in quantum thermodynamics \cite{Binder2018,Deffner2019} which is a rapidly developing field that explores the impact of 
quantum physics on thermodynamic laws and processes.  
The energy eigenstates and their mixtures form a natural set of incoherent states. In this context, the ability of quantum systems to be prepared in a superposition of energy eigenstates is of particular interest and importance, 
and it turns out that quantum coherence can potentially improve the performance of quantum thermodynamic schemes and machines \cite{Lostaglio2015a,Korzekwa2016,Narasimhachar2015}. 
During recent years, the resource theory of quantum coherence has been firmly established \cite{Streltsov2017,Chitambar2019}, and quantum coherence transformations and distillation by incoherent operations 
has been widely studied theoretically \cite{Du2015,Du2015B,Chitambar2016B,Regula2018,Fang2018,Chitambar2016,Liu2019,Liu2020,Fang2020} and tested experimentally \cite{Wu2017,Wu2018,Wu2020}.

Similarly to entanglement concentration and distillation \cite{Bennett1996B,Kwiat2001,Bennett1996,Deutsch1996,Pan2003,Hage2010}, probabilistic quantum operations can be useful for manipulation and conditional enhancement of quantum coherence \cite{Du2015,Chitambar2016B,Fang2018,Liu2020}. 
Probabilistic quantum operations can be viewed as nondestructive partial quantum measurements that conditionally apply a suitable quantum filter to the input quantum system. 
Here we focus on diagonal quantum filters that map incoherent states onto incoherent states and therefore belong to the class of strictly incoherent operations. 
Such quantum filter\footnote{This terminology is motivated by the fact that the quantum filter selectively attenuates amplitudes of certain basis states. 
	The quantum filters considered in the present work should be distinguished from the quantum filters that serve for estimation of quantum state parameters from a series of quantum measurements \cite{Belavkin1983,Belavkin1989,Belavkin1992}.}
is a trace-decreasing completely positive map $\hat{\rho}\rightarrow \hat{M}\hat{\rho}\hat{M}^\dagger$ 
described by a single Kraus operator $\hat{M}$ that is diagonal in the basis of incoherent states and satisfies $\hat{M}^\dagger\hat{M}\leq \hat{I}$.
Consequently, non-vanishing initial coherence is a necessary prerequisite for successful increase of coherence. 

In the context of quantum thermodynamics, such filters preserve energy eigenstates, but can conditionally change the mean energy of a state that is a mixture or superposition of several energy eigenstates.
Production of output states with increased coherence, resulting from the filtration process, can represent a useful resource in quantum thermodynamics. States with enhanced quantum coherence can speed up the energy transfer \cite{Farre2020, Nimmrichter2021} to the system of interest, 
or increase the power output when used as a coherent working medium in quantum thermal machines \cite{Klatzow2019}. If the filtered state is produced in larger number of copies, it may serve as improved coherent source of energy (fuel) for more efficient quantum machines \cite{Kurizki2016, Kurizki2018}. 
Input states formed by several subsystems such as $N$ two-level systems \cite{Gumberidze2019} are of particular interest, because the optimal enhancement of coherence may require filtering that introduces quantum correlations between the elementary subsystems.

Here, we further investigate the optimal conditional enhancement of quantum coherence and focus on optimal quantum filters that maximize the coherence of the output state for a given probability of success. We employ a linear optical setup to 
experimentally test the optimal conditional enhancement of coherence for a pair of two-level quantum systems represented by polarization states of single photons.
We experimentally demonstrate that collective quantum filters are optimal and yield better trade-offs between the achieved output coherence and the success probability than products of single-qubit quantum filters acting independently 
on each elementary two-level system. 

The rest of the paper is organized as follows. In Section II we derive the optimal quantum filters that maximize the coherence or energy of the filtered states for a fixed success probability of filtering. Analytical results are obtained for both pure and mixed initial states. 
The experimental setup is described in Section III and the experimental results are reported and discussed in Section IV. Finally, Section V presents a brief summary of our findings. Several technical details are collected in two Appendices.
Appendix A contains formal proof of properties of iterative coherence enhancement schemes inspired by iterative entanglement distillation protocols, and Appendix B includes details of comprehensive experimental characterization of the implemented two-qubit linear optical quantum filters. 

\section{Optimal quantum filters} 
Consider a pure input quantum state of a general $d$-level quantum system,
\begin{equation}
|\psi\rangle=\sum_{j=1}^d c_j|j\rangle,
\label{psidefinition}
\end{equation}
where $|j\rangle$ denotes the basis of incoherent states. In the context of quantum thermodynamics, we can assume this to be the basis formed by the energy eigenstates, $\hat{H} |j\rangle=E_j|j\rangle$. 
Throughout the manuscript, we keep this energetic viewpoint and assume that the energy levels are ordered in a nondecreasing manner, $E_j\leq E_k$ if $j\leq k$. 
Specifically, the input may be formed by $N$ two-level systems that are initially in a factorized state $|\psi\rangle=|\phi\rangle^{\otimes N}$, but our subsequent considerations are not restricted to this class of states.
We attempt to conditionally increase the mean energy and coherence of the state by partial quantum measurement, that conditionally applies to the input state the following quantum filter $\hat{M}$ diagonal in the energy basis,
\begin{equation}
\hat{M}=\sum_j m_j|j\rangle\langle j|, 
\label{Mdefinition}
\end{equation}
where $|m_j|^2\leq 1$. Such filtering can be implemented by suitable coupling to an auxiliary quantum system followed by measurement of the auxiliary system and conditioning on observation of certain measurement outcome.
For a general mixed state $\rho$ we define the mean energy
\begin{equation}
\bar{E}=\mathrm{Tr}[\hat{H} \hat{\rho}]=\sum_j E_j \rho_{jj},
\label{Emean}
\end{equation}
and the coherence \cite{Baumgratz2014}
\begin{equation}
C=\mathrm{Tr}[\hat{\rho}\log\hat{\rho}]-\mathrm{Tr}[\hat{\rho}_D\log\hat{\rho}_D],
\label{Cdefinition}
\end{equation}
where 
\begin{equation}
\hat{\rho}_D=\sum_{j}\rho_{jj}|j\rangle\langle j|,
\label{rhoDdefinition}
\end{equation}
is obtained from $\rho$ by a decoherence process that eliminates all off-diagonal density matrix elements and preserves the population of energy levels. 
After quantum filtering of $|\psi\rangle $ the normalized pure output state reads 
\begin{equation}
|\psi_{\mathrm{out}}\rangle=\frac{1}{\sqrt{P_S}}\hat{M}|\psi\rangle=\frac{1}{\sqrt{P_S}}\sum_{j}m_j c_j|j\rangle,
\end{equation}
where 
\begin{equation}
P_S=\langle \psi|\hat{M}^\dagger \hat{M}|\psi\rangle=\sum_j |m_j|^2 |c_j|^2 
\label{PSdefinition}
\end{equation}
denotes the probability of success of the filtering.

For further comparison with the enhancement of quantum coherence, let us first investigate what is the maximum achievable mean energy $\bar{E}$ for a given success probability $P_S$. We have
\begin{equation}
\bar{E}_{\mathrm{out}}= \frac{1}{P_S} \sum_j  E_j |m_j|^2 |c_j|^2.
\end{equation}
Since  $P_S$ is a fixed quantity representing the optimization constraint, we can 
equivalently maximize $P_S \bar{E}_{\mathrm{out}}$ instead of $\bar{E}_{\mathrm{out}}$, which simplifies the calculations. The resulting optimization problem can be formulated as maximization of 
\begin{equation}
Q_E=P_S \bar{E}_{\mathrm{out}}-\lambda P_S,
\label{QEdefinition}
\end{equation}
where $\lambda$ is a Lagrange multiplier. We simplify the notation by introducing $p_j=\rho_{jj}=|c_j|^2$ and $M_j=|m_j^2|$. With these definitions, we can write $Q_E$ as
\begin{equation}
Q_E=\sum_{j} M_j E_j p_j - \lambda \sum M_j p_j.
\end{equation}
Each $M_j$ must either satisfy the extremality condition 
\begin{equation}
\frac{\partial Q_E }{\partial M_k}=0,
\end{equation}
or lie at the boundary of the allowed values of $M_j$, i.e. $M_j=1$ or $M_j=0$. The extremality condition yields
\begin{equation}
(E_k-\lambda)p_k=0.
\label{extremalE}
\end{equation}
For nonvanishing $p_j$, Eq. (\ref{extremalE}) can be satisfied only if $\lambda$ is equal to one of the energy eigenvalues, $\lambda=E_j$.
We thus find that the optimal filtering consists of filters where the coefficients $M_{j}$ corresponding to energy level (or several degenerate energy levels) with energy $E_j$ 
can have arbitrary values in the allowed interval $[0,1]$ while all other coefficients are either equal to $0$ or $1$. Since our goal is to maximize $\bar{E}_{\mathrm{out}}$, the globally optimal strategy 
is to gradually eliminate and filter out the 
lowest energy eigenstates. Mathematically, the optimal filters consist of $d-1$ classes given by $m_j=0$, $j<k$, $m_j=1$, $j>k$, and $m_k \in [0,1]$, where $1 \leq k<d$ labels the filter. 
The values of $k$ and $m_k$ determine the success probability of filtering $P_S$ and vice versa.
In case of degenerate energy levels we can impose further symmetry and require that all $m_k$ corresponding to the same energy are equal. Note that the above derived filters are optimal also for general mixed input states because the assumption of state purity was not used in our calculations.

\begin{figure*}[t]
	\includegraphics[width=0.85\linewidth]{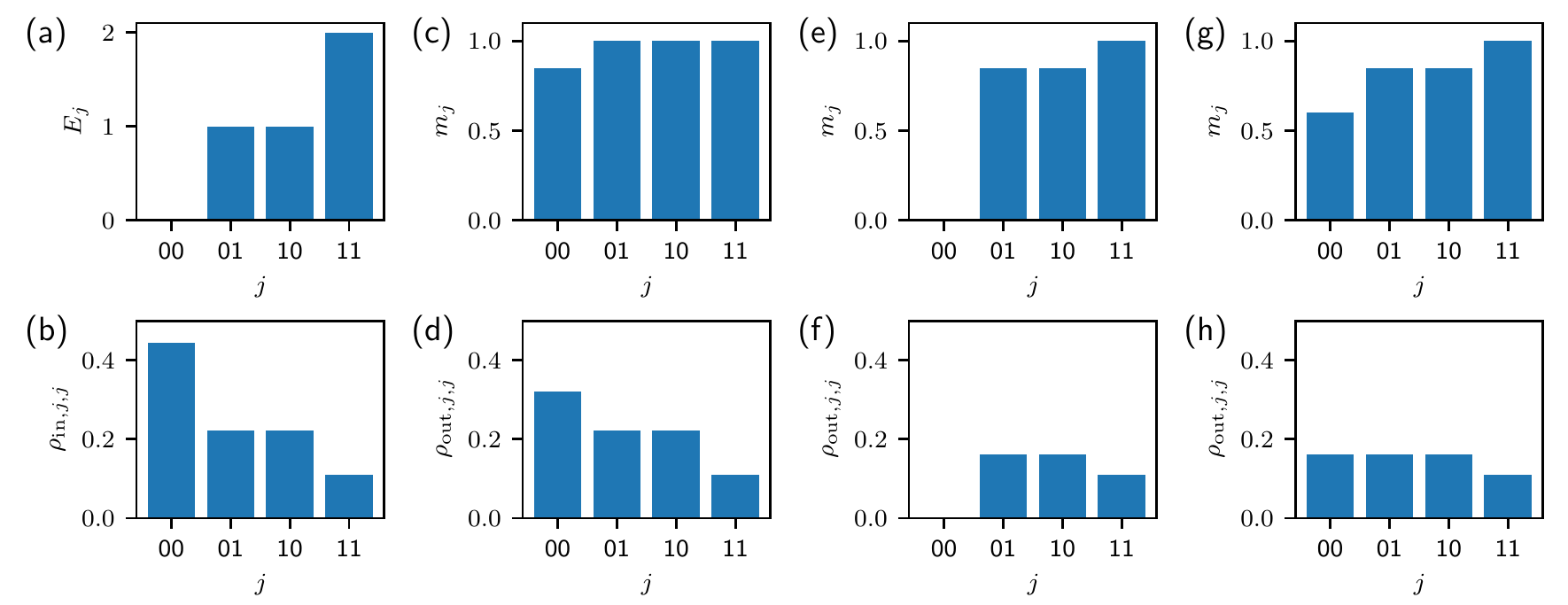}
	\caption{Example of optimal quantum filters for input two-qubit product state $|\phi\rangle|\phi\rangle$ with $p=1/3$, c.f. Eq. (\ref{psip}). Panels (a) and (b) show the energy eigenvalues and the initial state population, respectively. 
		Three examples of optimal quantum filters are given in panels (c,e,g), and the resulting non-normalized state populations are displayed in the corresponding bottom panels (d,f,h).}
\end{figure*}

Let us now turn our attention to the maximization of output coherence. For pure states, $\hat{\rho}=|\psi\rangle\langle \psi|$, we have $\mathrm{Tr}[\hat{\rho} \log \hat{\rho}]=0$ 
and the maximization of $C_{\mathrm{out}}$ for a given $P_S$ simplifies to maximization of
the entropy of $\hat{\rho}_{D,\mathrm{out}}$. We can further simplify the calculations by considering the equivalent maximization of 
\begin{equation}
Q_{C}=-P_S \mathrm{Tr}[\hat{\rho}_{D,\mathrm{out}}\log \hat{\rho}_{D,\mathrm{out}}+\log P_S] -\lambda P_S,
\end{equation}
where the entropy of $\hat{\rho}_{D,\mathrm{out}}$ was rescaled by a constant factor $P_S$ and a constant term $-P_S\log P_S$ was added. 
After some algebra, we get
\begin{equation}
Q_C= -\sum_{j}M_j p_j \log(M_j p_j)-\lambda \sum_j M_j p_j.
\end{equation}
The extremal conditions become
\begin{equation}
\frac{\partial Q_C}{\partial M_k}=-p_k\log(M_k p_k)-p_k-\lambda p_k=0.
\label{QCextremal}
\end{equation}
This yields
\begin{equation}
M_k=\frac{K}{p_k},
\label{MoptimalC}
\end{equation}
where $K$ is a normalization constant. We recall that we should also consider the extremal points $M_k=1$ or $M_k=0$. Since we aim at the maximization of coherence, 
$M_k=0$ can be ruled out as complete elimination of any component of the input state is not suitable for coherence maximization. 
By considering the condition (\ref{QCextremal}) and the properties of the entropy function $-x\log x$, whose first derivative is a decreasing function of $x$, 
we find that the optimal strategy is to subsequently attenuate all dominant amplitudes 
of the input quantum state until all the amplitudes are equal. To be more specific, assume that 
$p_j \geq p_k$ if $j\leq k$ and choose index $n$ such that $p_j> p_n$ if $j< n$. Then a subclass of the optimal filters reads
\begin{eqnarray}
M_j=\frac{K}{p_j}, \quad j<n, \nonumber \\
M_j=1, \quad j\geq n,
\label{MCoptimalX}
\end{eqnarray}
where $K\in [p_n,p_{n-1}]$. By considering all relevant $n$ we obtain the full set of optimal quantum filters that for a given success probability $P_S$ maximize the coherence of the output state.
The global optimality of the filters (\ref{MCoptimalX}) can be proved by showing that any filter that is not of the form (\ref{MCoptimalX}) is not optimal. Let $q_j=p_jM_j/P_S$ denote the normalized probabilities of the basis states $|j\rangle$ after successful filtering.
Choose $l$ such that $q_l\geq q_j$, $\forall j$. Assume that there exists $k$ such that $M_k < 1$ and $q_k<q_l$. Then one can define a different filter 
with the same $P_S$ that would yield $\tilde{q}_l=q_l-dq$ and $\tilde{q}_k=q_k+dq$, with $dq>0$ being an infinitesimal probability change, while all other $q_j$ remain unmodified.
This modified filter achieves higher output coherence for a given $P_S$, because 
\[
dC=-\tilde{q}_l\log\tilde{q}_l-\tilde{q}_k\log\tilde{q}_k+q_l\log q_l+q_k\log q_k
\]
is positive, $dC=dq \log(q_l/q_k)>0$. This proves that a quantum filter that would actively reduce some output probability $q_k$ below the largest output probability $q_l$ by active filtering (i.e. $M_k<1$), cannot be optimal.

These analytical results can not be straightforwardly extended to mixed states, for which the term $\mathrm{Tr}\hat{\rho}\log\hat{\rho}$ does not vanish. However, we can obtain analytical formulas for the optimal quantum filters 
even for mixed states, if we consider different quantification of coherence based on Tsallis entropy
\begin{equation}
S_2(\rho)=1-\mathrm{Tr}(\hat{\rho}^2).
\end{equation}
In particular, we can define
\begin{equation}
\tilde{C}=S_2(\hat{\rho}_D)-S_2(\hat{\rho})=\mathrm{Tr}(\hat{\rho}^2)-\mathrm{Tr}(\hat{\rho}_D^2).
\end{equation}
For this quantification of coherence, we can find the optimal quantum filter that maximizes $\tilde{C}$ for a given $P_S$ by maximizing the function
\begin{equation}
\tilde{Q}_C=P_S^2 \tilde{C}- \lambda P_S= P_S^2[\mathrm{Tr}(\hat{\rho}^2)-\mathrm{Tr}(\hat{\rho}_D^2)]-\lambda P_S,
\end{equation}
that can be rewritten as
\begin{equation}
\tilde{Q}_C=\sum_{j,k} M_j M_k \rho_{jk}\rho_{kj} -\sum_{j} M_j^2 \rho_{jj}^2 - \lambda \sum_{j} M_j \rho_{jj}.
\end{equation}
The extremality conditions 
\begin{equation}
\frac{\partial \tilde{Q}_C}{\partial M_j} =2\sum_{k} |\rho_{jk}|^2 M_k-2\rho_{jj}^2 M_j- \lambda \rho_{jj}=0,
\label{Cextremalmixed}
\end{equation}
form a system of linear equations that can be solved to determine $M_k$. Note that one also has to consider the extremal points $M_j=1$ 
and $M_j=0$ and optimize over all combinations of these extremal points for some $M_j$ and solutions 
of the system of extremal equations (\ref{Cextremalmixed}) for the remaining $M_k$. For pure states, $|\rho_{jk}|^2=\rho_{jj} \rho_{kk}$, and from Eq. (\ref{Cextremalmixed}) we recover the optimality conditions (\ref{MoptimalC}).

Let us illustrate the optimal filtering procedures for coherence or energy enhancement 
on the example of a pair of two-level systems prepared in a pure state $|\phi\rangle|\phi\rangle$, where $|\phi\rangle=\sqrt{1-p}|0\rangle+\sqrt{p}|1\rangle$ with $p<0.5$. 
Written explicitly, the state reads
\begin{equation}
|\psi\rangle=(1-p)|00\rangle+\sqrt{p(1-p)}(|01\rangle+|10\rangle)+p|11\rangle.
\label{psip}
\end{equation}
Here $|0\rangle$ and $|1\rangle$ denote the ground and excited energy eigenstate of each two-level system, with energy difference $\Delta E$. The energy spectrum of the system is plotted in Fig. 1(a). The filters that optimize
the trade-off between the mean energy of the system (\ref{Emean}) and the success probability of filtering (\ref{PSdefinition}) are given by
\begin{equation}
\hat{M}_{E,1}=\frac{\sqrt{P_S-P_{\mathrm{th}}}}{1-p}|00\rangle\langle 00|+|01\rangle\langle 01|+|10\rangle\langle 10|+|11\rangle\langle 11|
\label{ME1}
\end{equation}
if $P_S\geq P_{\mathrm{th}}$, where $P_{\mathrm{th}}=p(2-p)$, and
\begin{equation}
\hat{M}_{E,2}=\sqrt{\frac{P_S-p^2}{2p(1-p)}}(|01\rangle\langle 01|+|10\rangle\langle 10|)+|11\rangle\langle 11|, 
\label{ME2}
\end{equation}
if $p^2 \geq P_S<P_{\mathrm{th}}$.
Examples of the optimal filters (\ref{ME1}) and (\ref{ME2}) are given in Figs.~1(c,e) and their impact on specific input state with $p=1/3$ is illustrated in Fig.~1(d,f).
The filters that optimize the trade-off between the coherence (\ref{Cdefinition}) and the success probability (\ref{PSdefinition}) read
\begin{equation}
\hat{M}_{C,1}=\frac{\sqrt{P_S-P_{\mathrm{th}}}}{1-p}|00\rangle\langle 00|+|01\rangle\langle 01|+|10\rangle\langle 10|+|11\rangle\langle 11|, 
\label{MC1}
\end{equation}
if $P_S \geq P_{\mathrm{th}}+p(1-p),$ and
\begin{equation}
\hat{M}_{C,2}=b\sqrt{\frac{p}{1-p}}|00\rangle\langle 00|+b|01\rangle\langle 01|+b|10\rangle\langle 10|+|11\rangle\langle 11|,
\label{MC2}
\end{equation}
where 
\begin{equation}
b=\sqrt{\frac{P_S-p^2}{3p(1-p)}},
\end{equation}
and $4p^2 \geq P_S<P_{\mathrm{th}}+p(1-p)$.
Examples of optimal quantum filters (\ref{MC1}) and (\ref{MC2}) are presented in Figs. 1(c,g) with their impact on the input state illustrated in figs. 1(d,h). For states with non-vanishing initial coherence and low enough mean energy the 
quantum filtering can simultaneously increase both energy and coherence \cite{Gumberidze2019}. In particular, in the above example the filter $\hat{M}_{C,1}$ is optimal for both energy and coherence improvement.
However, at some point further increasing the energy leads to coherence reduction and vice versa. Moreover, if
the state with maximum energy is nondegenerate, then the maximum energy corresponds to zero coherence. 

For the sake of completeness, we also specify the optimal pure states $|\psi\rangle=\sum_j \sqrt{p_j}|j\rangle$ 
that for a given fixed energy $E$ exhibit maximum coherence, i.e. maximum entropy of the diagonal state $\rho_D$. 
This class of states can serve as a reference and benchmark when evaluating the performance of coherence and energy enhancement.
We recover the well-known result from statistical physics that $p_j$ is a thermal distribution and
\begin{equation}
|\psi\rangle= \frac{1}{\sqrt{Z(\beta)}}\sum_{j} e^{-\beta E_j/2}|j\rangle,
\end{equation}
where $Z(\beta)=\sum_j \exp(-\beta E_j)$. This directly follows from the fact that for a fixed mean energy $\bar{E}=\mathrm{Tr}[\hat{H}\hat{\rho}_D]$ the entropy of $\hat{\rho}_D$ is maximized if $\hat{\rho}_D=e^{-\beta \hat{H}}/Z(\beta)$ is a thermal state. 
Global maximum of $C$ is reached in the limit of infinite temperature, $\beta=0$, when $|\psi\rangle=\frac{1}{\sqrt{d}}\sum_{j=1}^d|j\rangle$ and $C=d\log d$.

In entanglement distillation, single-copy protocols based on quantum filtering \cite{Bennett1996B,Kwiat2001} may be outperformed by multicopy iterative entanglement distillation schemes \cite{Bennett1996,Deutsch1996,Pan2003,Hage2010}. 
However, this does not hold for the coherence enhancement investigated in the present work. Specifically, one can consider iterative coherence synthesis protocols where the quantum filters are applied to pairs of $d$-level quantum systems 
and one $d$-level system from each output pair forms an input for the next stage of the protocol.
We find that the resulting output state after $n$ iterations is formed by a mixture of states filtered with various filters diagonal in the energy basis. This indicates that the iterative procedure becomes equivalent to a mixture of single-copy filterings. 
Moreover, since all the involved quantum filters mutually commute, it turns out that the costly iterative procedure can be replaced by a 
more efficient and simpler scheme where a sequence of non-destructive mutually commuting generalized quantum measurements is applied to a single input pair of $d$-level systems. Technical proof of these findings is provided in Appendix A.

\section{Experimental setup}

We aim to test the measurement-based enhancement of quantum coherence with linear optics \cite{Kok2007,Kok2014,Flamini2019}, where 
qubits are represented by polarization states of single photons. Specifically, we focus on coherence enhancement of a pair of qubit systems initially prepared in a product state. For pure input 
states $|\phi\rangle|\phi\rangle$ the optimal quantum filters are given by Eqs. (\ref{ME1})-(\ref{MC2}) and they all belong to the class of filters 
\begin{equation}
\hat{M}=a|00\rangle 00|+b(|01\rangle \langle 01|+|10\rangle \langle 10|)+|11\rangle\langle 11|,
\label{MfilterLO}
\end{equation}
with $a\leq b^2$. For $a=b^2$ the two-qubit filter (\ref{MfilterLO}) factorizes into a product of two single-qubit filters $\hat{M}_S=b|0\rangle\langle 0| +|1\rangle\langle 1|.$
We note that the filters (\ref{MfilterLO}) can be utilized also for mixed product input two-qubit states $\hat{\rho}\otimes\hat{\rho}$ with $\rho_{00} > \rho_{11}$.

We implement the two-qubit quantum filters (\ref{MfilterLO}) for qubits encoded into states of single photons by interference of the two photons in a suitably designed optical interferometer, followed by postselection of cases when a single photon is present in each output of the filter. 
The linear optical filter thus operates in the coincidence basis, similarly to many linear optical quantum gates \cite{Kok2007,Okamoto2005,Langford2005,Kiesel2005}. The linear optical realization of the filter (\ref{MfilterLO}) 
can impose an extra cost in terms of reduced overall success probability of filtering \cite{Fiurasek2021}. 
This means that instead of filter $\hat{M}$ we implement an equivalent filter $\sqrt{P_L} \hat{M}$, where $P_L\leq 1$ is the probability reduction factor. 

\begin{figure}[t]
	\centerline{\includegraphics[width=\linewidth]{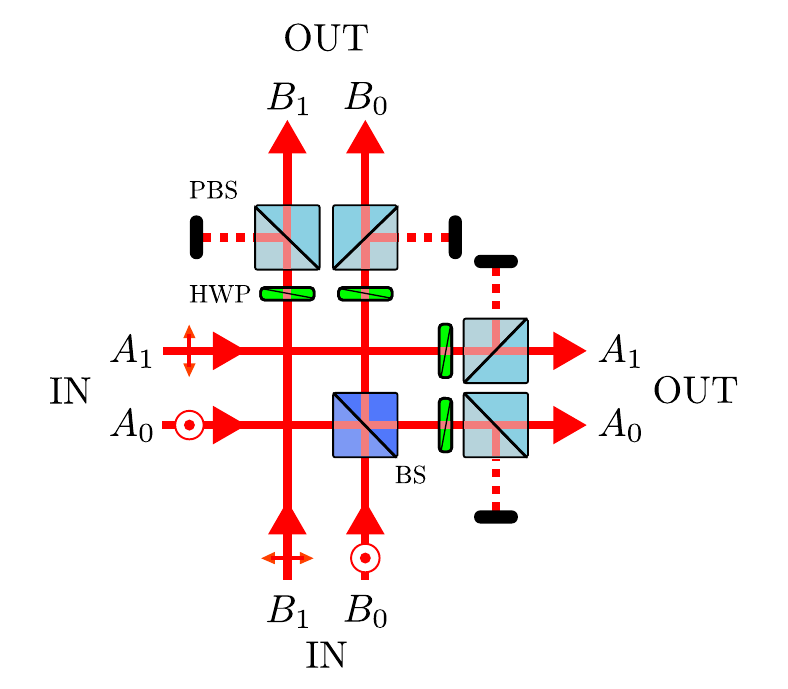}}
	\caption{Linear optical implementation of two-qubit quantum filters (\ref{MfilterLO}) with $a\leq b^2$. The modes $A_0$ and $B_0$ that encode the logical state $|0\rangle$ are coupled at a beam splitter BS with suitably chosen transmittance. 
		Subsequently, either modes $A_0$ and $B_0$ or modes $A_1$ and $B_1$ are attenuated by sending them through a sequence of suitably rotated half-wave plates HWP and polarizing beam splitters PBS 
		whose auxiliary input modes are prepared in vacuum state. Note that the input modes $A_0$ and $B_0$ are vertically polarized, while the modes $A_1$ and $B_1$ are horizontally polarized, which corresponds to the actual situation in our experiment.
	}
	\label{fig-interferometer}
\end{figure}

Let $A_0, A_1$ and $B_0,B_1$ denote the modes encoding the logical states $|0\rangle$ and $|1\rangle$ of qubits A and B, respectively.
Quantum filters (\ref{MfilterLO}) with $a\leq b^2$ can be implemented by interference of modes $A_0$ and $B_0$ 
on an unbalanced beam splitter, 
followed by attenuation of modes $A_0$ and $B_0$ or $A_1$ and $B_1$, 
depending on the values of $a$ and $b$ \cite{Fiurasek2021}. This interferometric scheme depicted in Fig.~\ref{fig-interferometer} is optimal and maximizes the probability factor $P_L$.
Practical implementation of this optimal setup requires a tunable beam splitter, that can be realized for instance
as a balanced Mach-Zehnder interferometer with phase shift in one arm controlling the transmittance. Alternatively, one can consider configuration 
where the two coupled modes are spatially overlapped and orthogonally polarized and are coupled via half-wave plate. In all configurations one needs to achieve sufficient interferometric stability.

\begin{figure}[t]
	\includegraphics[width=\linewidth]{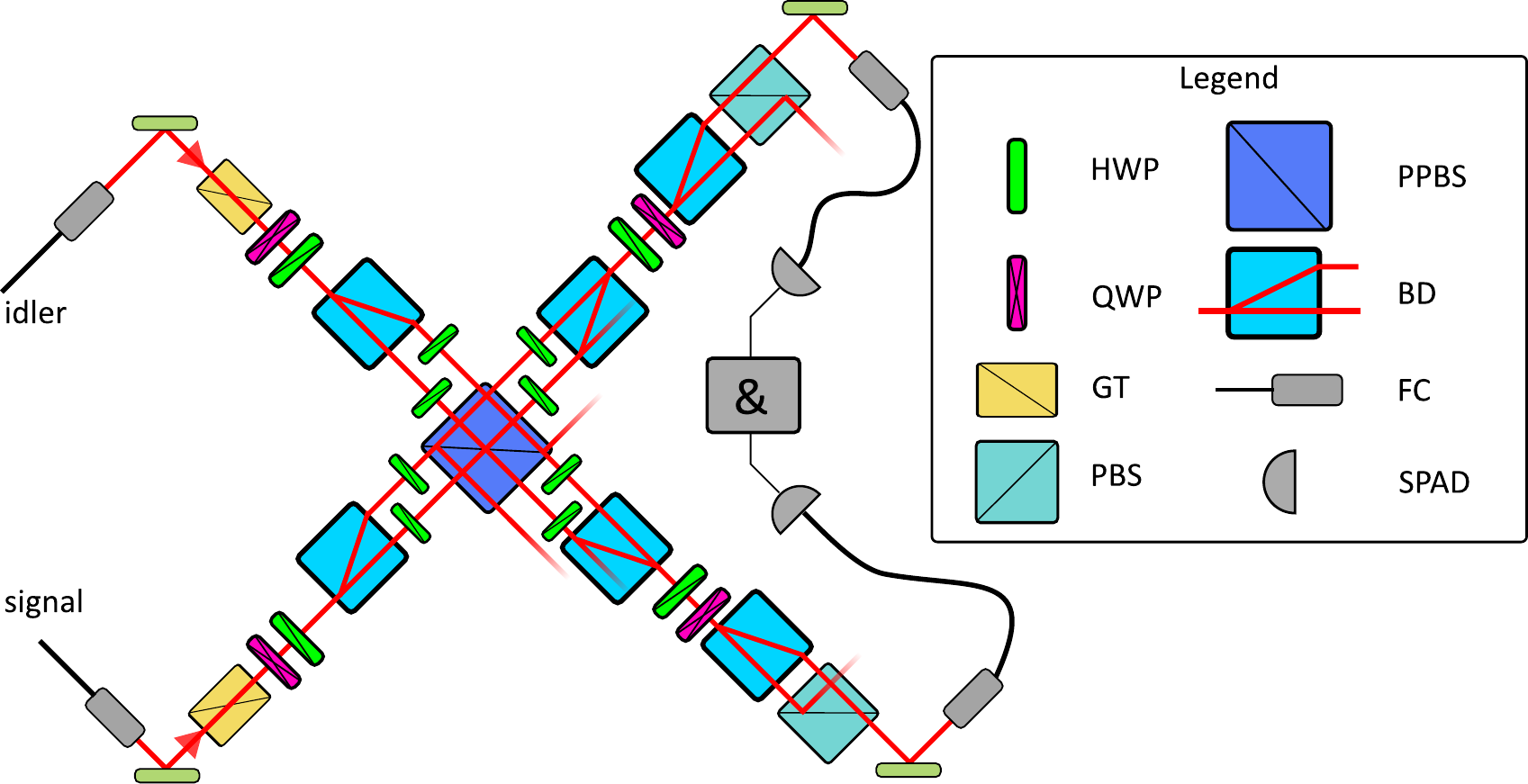}
	\caption{Experimental setup. HWP - half-wave plate, QWP - quarter-wave plate, GT - Glan-Taylor polarizer, PBS - polarizing beam splitter, PPBS - partially polarizing beam splitter, BD - calcite beam displacer, FC - fiber collimator, 
		SPAD - single-photon avalanche diode.}
	\label{fig-setup}
\end{figure}

To address these experimental issues, we have chosen to employ the configuration as shown in Fig.~\ref{fig-interferometer} but with a fixed beam splitter BS and additional half-wave plates inserted into the paths of the input modes. The complete setup, shown in Fig.~\ref{fig-setup}, involves two crossed inherently stable 
Mach-Zehnder interferometers formed by pairs of calcite beam displacers \cite{Lanyon2009,Starek2016B}. 
This inherently stable and compact setup provides sufficient flexibility to control the effective coupling between the modes $A_0$ and $B_0$ and emulate a beam splitter with arbitrary transmittance 
by suitable rotation of the waveplates in the setup \cite{Starek2016}. Although our scheme generally does not reach the maximum possible $P_L$, 
this is not a significant obstacle in our proof-of-principle experiment, where we achieve sufficient two-photon 
coincidence rates to collect enough data for full tomographic characterization of the investigated quantum filters and two-photon states at the filter output.

In our experiment, we generate time-correlated photon pairs with central wavelength of $810$ nm in type-II spontaneous parametric down-conversion process in a nonlinear BBO crystal pumped by a
100-mW 405 nm laser diode. The generated signal and idler photons are coupled into single-mode optical fibers and guided to the main experiment.
In the experimental setup, depicted in Fig.~\ref{fig-setup}, we decouple the photons from the fibers into collimated beams 
using 11-mm collimators and purify their polarization using Glan-Taylor prisms (GT). 
The prisms are oriented to prepare horizontal polarization, encoding the state $|1\rangle$. 
Pairs of quarter- and half-wave plates manipulate the photon's polarization state to prepare any desired input state.

Calcite beam-displacing crystals (BD) laterally shift horizontally-polarized photons while leaving the vertically-polarized photons intact. 
The first beam displacers thus split the input signals into two parallel beams with 6 mm spacing and introduce correlations between the photon polarization and propagation path. 
Two ordinary beams, corresponding to the originally vertically polarized components, interfere at the central partially-polarizing beam splitter (PPBS). 
The PPBS has transmittance $T_{H} = 1$ and $T_{V} = 1/3$ for horizontally and vertically polarized photons, respectively. 
The remaining beams do not overlap at the PPBS, and potential reflections are not detected. Each pair of parallel beams is subsequently recombined at the second beam displacer. 
The waveplates inserted inside the calcite interferometers control the attenuation and the effective strength of the interferometric coupling between the two photons.

Successful application of the two-qubit quantum filter (\ref{MfilterLO}) is indicated by coincidence detection of one photon at each output port of the setup. We can project each photon onto an arbitrary polarization state 
using a half-wave and quarter-wave plate followed by a calcite crystal and a polarizing beam splitter. Single-photon avalanche diodes detect the photons, and the resulting electronic signal is processed in coincidence electronics. 
The number of single detection events and the number of simultaneous detection events are recorded with a counter of electronic pulses.
The stability of the Mach-Zehnder interferometers formed by calcite beam displacers can be characterized by the standard deviation of fast phase fluctuations, which is approximately $2^\circ$. Moreover, the
phase also drifts around $1^\circ$ per hour when the setup is sufficiently isolated from the airflow in the laboratory. This high passive stability enabled us to carry out continuous measurements for several hours before readjusting the setup.

\section{Experimental results}

\begin{figure}[t]
	\centerline{\includegraphics[width=\linewidth]{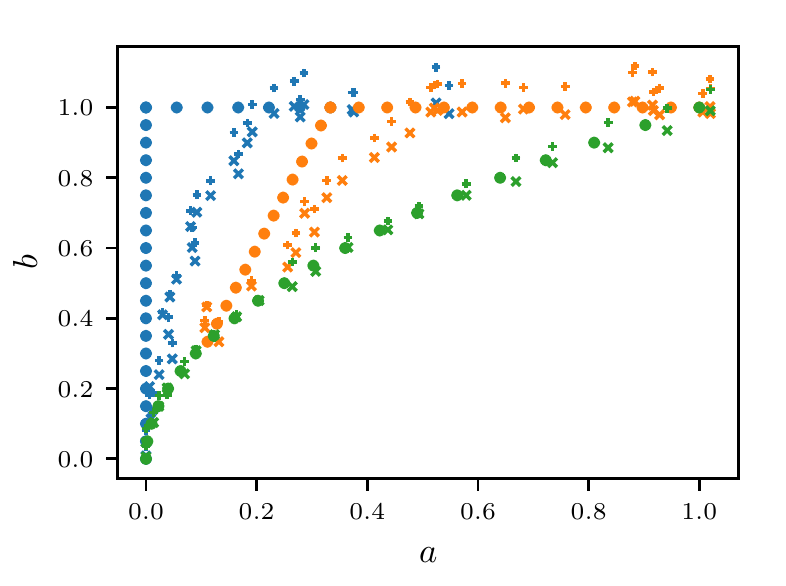}}
	\caption{Optimal symmetric quantum filters for enhancement of energy (blue) and coherence (orange) are shown for pure input two-qubit state $|\phi\rangle|\phi\rangle$ with $p=0.1$. 
		In the region $b=1$, $a>\sqrt{p/(1-p)}$ the filters optimal for energy and coherence enhancement coincide. 
		Dots indicate the optimal theoretical filters, crosses show the experimentally achieved filter parameters. Since the experimentally realized filters are not perfectly symmetric, 
		we plot two values of $b$ for each filter, one for state $|01\rangle$ ($\times$), and the other for state $|10\rangle$ ($+$). For reference, in green we plot the factorized filters characterized by relation $a=b^2$. 
		These filters are products of two identical diagonal single-qubit filters that attenuate state $|0\rangle$. Statistical error bars are smaller than the marker size and therefore not plotted.}
	\label{fig-filterparameters}
\end{figure}

We have comprehensively characterized the implemented linear optical two-qubit quantum filters (\ref{MfilterLO}) by quantum process tomography. The observed quantum process fidelities of the filters exceeded $0.96$ and the quantum process purities exceeded $0.95$, which indicates 
very good quality of the implemented filters. For details, see Appendix B. We utilize the quantum filters to improve the overall coherence and total energy of a pair of two-level quantum systems. We first consider pure input state $|\phi\rangle|\phi\rangle$ with 
\begin{equation}
|\phi\rangle=\sqrt{1-p}|0\rangle+\sqrt{p}|1\rangle.
\end{equation}
The optimal filters that for a given probability of success $P_{S}$ maximize the mean energy or coherence were specified in Section II, c.f. Eqs. (\ref{ME1})-(\ref{MC2}). The optimal filters for $p=0.1$ are further illustrated in Fig.~\ref{fig-filterparameters} 
as dots in the $a-b$ parametric space, where the orange and blue colors encode filters optimal for coherence and energy enhancement, respectively. Initially, it is optimal to use filters (\ref{MfilterLO}) with $b=1$ and $a<1$ until $a=\sqrt{p/(1-p)}$ is reached, 
and this filtering is simultaneously optimal for both the coherence and energy enhancement. 
For $a<\sqrt{p/(1-p)}$ the two optimal strategies separate. Optimal coherence enhancement requires filters with $b<1$ and $a=b\sqrt{p/(1-p)}$ until the point $a=b^2=p/(1-p)$ is reached. On the other hand, for the energy enhancement it is optimal to first reduce 
the amplitude of the state $|00\rangle$ to $0$ and only then decrease the amplitudes of the states $|01\rangle$ and $|10\rangle$ by filters with $a=0$ and $b<1$. 
For comparison, in green we also plot the symmetric filters satisfying $a=b^2$, that can be implemented by independent filtering of each input qubit,
$|\phi\rangle \rightarrow b\sqrt{1-p}|0\rangle+\sqrt{p}|1\rangle$.

In order to check the filters we perform auxiliary measurements with input basis states $|jk\rangle$. 
The actual values of $a$ and $b$ determined from these measurements are plotted in Fig.~\ref{fig-filterparameters} as crosses. 
Note that for each value of $a$ two crosses are displayed, one corresponding to attenuation of state $|01\rangle$ and the other to attenuation of state $|10\rangle$. 
Although the filters are nominally symmetric, in practice small discrepancies between the attenuation of the states $|01\rangle$ and $|10\rangle$ occur. 
Several values of $b$ slightly exceed $1$, which is caused by the chosen normalization that the amplitude of $|11\rangle$ remains unchanged. However, several experimentally 
implemented filters slightly attenuate the state $|11\rangle$ with respect to $|01\rangle$ and $|10\rangle$ which formally results in $b>1$.
The factorized filters are implemented with high precision, see the green dots and crosses. The optimal entangling filters are more difficult to implement and 
the largest discrepancies occur for filters close to $a=0$ and $b=1$, which requires perfect destructive quantum interference to completely eliminate the component $|00\rangle$ of the state.
Note that the parameters $a$ and $b$ are attenuation amplitudes and therefore for example $a=0.1$ means that the population of level $|00\rangle$ is attenuated by factor $a^2=0.01$.

\begin{figure}
	\centerline{\includegraphics[width=\linewidth]{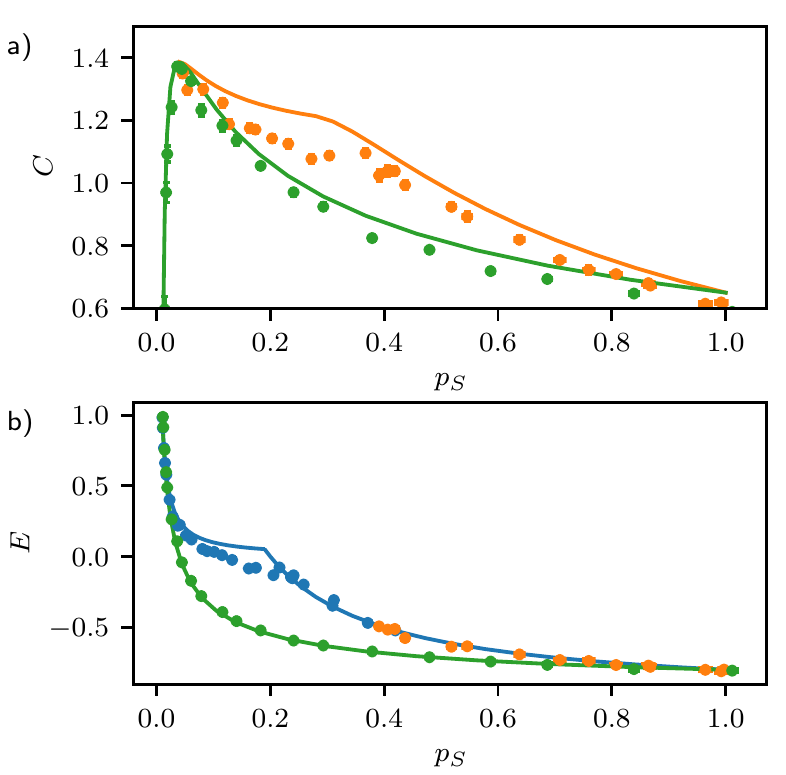}}
	\caption{Dependence of coherence $C$ (a) and mean energy $\bar{E}$ (b) of the output filtered state on the probability of successful filtering $P_S$ is plotted for pure input state $|\phi\rangle|\phi\rangle$ with $p=0.1$. Dots show experimental data
		and the lines represent theoretical predictions. 
		Blue and orange dots and curves represent results for the optimal filters (c.f. Fig.~\ref{fig-filterparameters}), and the green dots and curves indicate results for factorized filters with $a=b^2$.
		For most data, error bars are smaller than the symbol size.}
	\label{fig-pureresults}
\end{figure}

\begin{figure*}[t!]
	\includegraphics[width=0.8\linewidth]{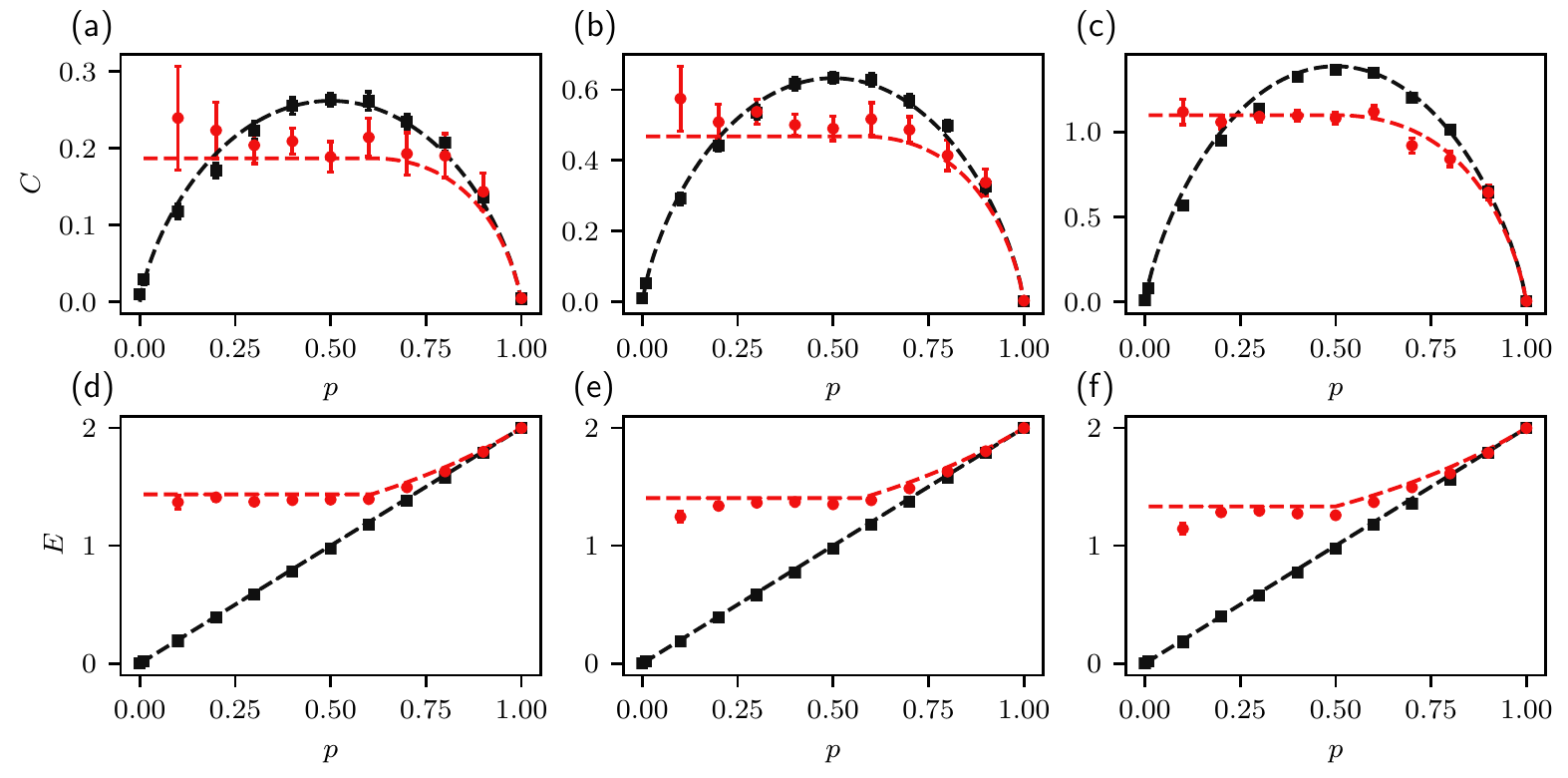}
	\caption{Experimental filtering of products of two mixed states (\ref{rhomixed}). Filters with $a=0$ are utilized and the parameter $b$ is optimized for each input state to maximize the output coherence. 
		The output coherence $C$ and mean energy $\bar{E}$ are plotted as functions of the initial state population $p$ for three different degrees of mixedness $\eta=0.5$ (a,d), $\eta=0.75$ (b,e), and $\eta=1$ (c,f). 
		Red dots represent experimental data and red lines indicate the theoretical predictions. For comparison, the black lines and dots show the energy and coherence of the input states.}
	\label{fig-mixed}
\end{figure*}

We have applied the optimal quantum filters shown in Fig.~\ref{fig-filterparameters} to the input state $|\phi\rangle|\phi\rangle$ with $p=0.1$ and we have performed full tomography of the output two-qubit quantum states. 
From the recorded data and the auxiliary reference measurements on input basis states $|jk\rangle$ we have determined the factor $P_L$ and the inherent probability of successful quantum filtering $P_S=P_T/P_L$, 
where $P_T$ is the total experimentally observed success probability. When estimating $P_S$ we thus compensate for the extra reduction of the success probability imposed by our linear optical implementation of the quantum filters.
The experimental results are shown in Fig.~\ref{fig-pureresults} where we plot the dependence of coherence and mean energy of the output states on the success probability $P_S$. 
For comparison, we plot also the results obtained for factorized filters applied independently to each two-level system.
We can see that the optimal filters clearly outperform the factorized filters and the experimental results are in good agreement with theory. The discrepancies between theory and experiment are larger for the enhancement of coherence, which 
is sensitive to preservation of purity of the quantum state, while the energy solely depends on the state populations, i.e. diagonal density matrix elements in the energy basis.

We have also tested filtering of input mixed states. Specifically we have considered input product states $\hat{\rho}\otimes \hat{\rho}$, where 
\begin{equation}
\hat{\rho}= \left(
\begin{array}{cc}
1-p & \eta \sqrt{p(1-p)} \\
\eta \sqrt{p(1-p)} & p
\end{array}
\right).
\label{rhomixed}
\end{equation}
The parameter $\eta$ controls state purity, $\mathrm{Tr}\hat{\rho}^2=1-2\eta p(1-p)$. We have experimentally generated the mixed states $\hat{\rho}$ as statistical mixtures of pure states. 
We have tested one of the scenarios considered in Ref. \cite{Gumberidze2019}, where the lowest energy state is completely filtered out, $a=0$, and $b$ is optimized
to maximize the output state coherence. Note that these filters belong to the class of filters optimal for the energy enhancement, c.f. Eq. (\ref{ME2}). Numerical calculations reveal that there exists a threshold $p_{\mathrm{th}} \geq 0.5$ that depends on $\eta$. 
For $p\leq p_{\mathrm{th}}$ the optimization yields constant $C$ and $\bar{E}$
that depend only on $\eta$ but not on $p$, see the theoretical curves plotted in Fig.~\ref{fig-mixed}. For $p>p_{\mathrm{th}}$ it is optimal to set $b=1$ within the considered class of filters and the coherence decreases with increasing $p$ while the mean energy increases.
The measurements were performed for a range of different input values of $p$ and for three different degrees of mixedness $\eta$. 
The experimental results are reported in Fig.~\ref{fig-mixed} and are in very good agreement with the theoretical predictions. We can observe that for low $p$ this choice of filters can simultaneously increase both the energy and coherence. 

\section{Summary}
In summary, we have theoretically and experimentally investigated enhancement of overall quantum coherence and mean energy of quantum states by quantum filters diagonal in the energy basis. 
Such quantum filtering can serve for synthesis of quantum coherence, where several weakly coherent two-level quantum systems with low energy are synthesized 
into a single large quantum system exhibiting improved coherence and increased energy. 
We have presented the optimal quantum filters that for a given success probability of filtering maximize either the output energy or coherence. 
We have shown that analytical results for enhancement of coherence can be obtained even for input mixed states provided that the coherence is quantified 
by Tsallis entropy based on state purity. We have also analyzed iterative measurement-based synthesis protocol for pairs of quantum systems and we have proved 
that it does not bring any advantage because all the applied quantum filters mutually commute. 
We have tested and verified the performance of the optimal quantum filters with linear optics setup, 
where a pair of two-level quantum systems is represented by polarization states of two photons and the two-photon
filtering is implemented with a suitable optical interferometer followed by postselection on coincidence detection of a single photon in each output port of the filter. 
We have experimentally confirmed the superiority of collective quantum filters that outperform factorized products of single-qubit 
filters and for given success probability of filtering yield higher output coherence or energy.

\acknowledgments
We acknowledge support by the Czech Science Foundation under Grant No. 19-19189S. R. S. acknowledges support from Palack\'{y} University under Grant. No. IGA-PrF-2021-006.

\appendix

\section{Analysis of iterative coherence enhancement}

Here we prove that iterative protocols based on partial quantum measurements diagonal in the energy basis are not advantageous for the enhancement of quantum coherence and mean energy.
We consider protocol where the quantum filter $\hat{M}$ is jointly applied to a pair of $d$-dimensional systems A and B prepared initially in uncorrelated quantum states $\hat{\rho}$. The total input state reads $\hat{\rho}_A \otimes\hat{\rho}_B$, and the quantum 
filter diagonal in the energy basis $|ij\rangle$ can be written as
\begin{equation}
\hat{M}=\sum_{i,j=1}^d m_{ij} |ij\rangle \langle ij|.
\end{equation}
The conditional quantum state of system A after the successful filtering reads
\begin{equation}
\hat{\rho}_{\mathrm{out},A}=\frac{1}{P_S}\sum_{j=1}^d \rho_{jj} \hat{K}_j \hat{\rho} \hat{K}_j^\dagger,
\end{equation}
where
\begin{equation}
\hat{K}_j=\sum_{k=1}^d m_{kj}|k\rangle\langle k|.
\end{equation}
We can see that the output state $\hat{\rho}_{\mathrm{out},A}$ is a mixture of input states filtered with operators $\hat{K}_j$ diagonal in the energy basis. 
Suppose now that we will use two copies of the output state $\hat{\rho}_{\mathrm{out},A}$ as an input state 
for the next stage of the iterative protocol, with possibly different quantum filter $\hat{M}'$. The full non-normalized conditional output state of systems A and B will be given by
\begin{equation}
\hat{\sigma}_{AB}=\sum_{j,k=1}^d  \rho_{jj}\rho_{kk} \hat{M}' (\hat{ K}_j\otimes \hat{K}_k) ( \hat{\rho}\otimes \hat{\rho} )(\hat{K}_j^\dagger\otimes \hat{K}_k^\dagger)  \hat{M}'^\dagger .
\label{sigmaAB}
\end{equation}
Since all operators $\hat{M}'$ and $\hat{K}_j$ are diagonal in the energy basis, we can rewrite Eq. (\ref{sigmaAB}) in a more compact form
\begin{equation}
\hat{\sigma}_{AB}=
\sum_{j,k=1}^d \hat{W}_{jk}  \hat{\rho}\otimes\hat{\rho} \hat{W}_{jk}^\dagger,
\label{sigmaABcompact}
\end{equation}
where $\hat{W}_{jk}=\sqrt{\rho_{jj}\rho_{kk}}\hat{M}' (\hat{K}_j\otimes \hat{K}_k)$ are operators diagonal in the energy basis. The output state $\hat{\sigma}_{AB}$ is thus a mixture of output states obtained from the initial input 
$\hat{\rho}\otimes\hat{\rho}$ by application of diagonal filter $\hat{W}_{jk}$. The Kraus operators $\hat{W}_{jk}$ specify a trace-decreasing completely positive map and 
\begin{equation}
\sum_{j,k} \hat{W}_{jk}^\dagger \hat{W}_{jk} \leq \hat{I}
\label{CPmapcondition}
\end{equation}
by construction. For convenience, we switch to single-index labeling $\hat{W}_l$, $l=1,\ldots,d^2$. Since all operators $\hat{W}_{l}$ are diagonal in the energy basis and commute, 
the transformation (\ref{sigmaABcompact}) can be implemented by sequence of non-destructive two-element generalized quantum measurements, where each measurement applies one of two complementary filters $\hat{M}_{+,l}$ or $\hat{M}_{-,l}$ given by 
\begin{eqnarray}
\hat{M}_{+,l}&=&\hat{W}_l \left(I-\sum_{m=1}^{l-1} \hat{W}_m^\dagger \hat{W}_m\right)^{-1/2} \nonumber \\
\hat{M}_{-,l}&=&\left(\hat{I}- \hat{M}_{+,l}^\dagger \hat{M}_{+,l}\right)^{1/2}. 
\label{POVMMl}
\end{eqnarray}
The matrix inverse in the expression for $\hat{M}_{+,l}$ is the Moore-Penrose pseudoinverse.
To verify that the operators (\ref{POVMMl}) describe valid quantum measurements we need to show that $\hat{M}_{+,l}^\dagger \hat{M}_{+,l}\leq \hat{I}$. This follows from the inequality 
\begin{equation}
\hat{ W}_{l}^\dagger \hat{W}_l \leq \hat{I}-\sum_{m=1}^{l-1} \hat{W}_{m}^\dagger \hat{W}_m,
\end{equation}
which is an immediate consequence of the condition (\ref{CPmapcondition}). The measurements are applied sequentially until a plus outcome $\hat{M}_{+,j}$ is obtained, 
at which point the measurement sequence is stopped and an output state is produced. If all measurements yield the minus outcome $\hat{M}_{-,j}$, then the filtering was unsuccessful.

\begin{figure}[b]
	\centerline{\includegraphics[width=\linewidth]{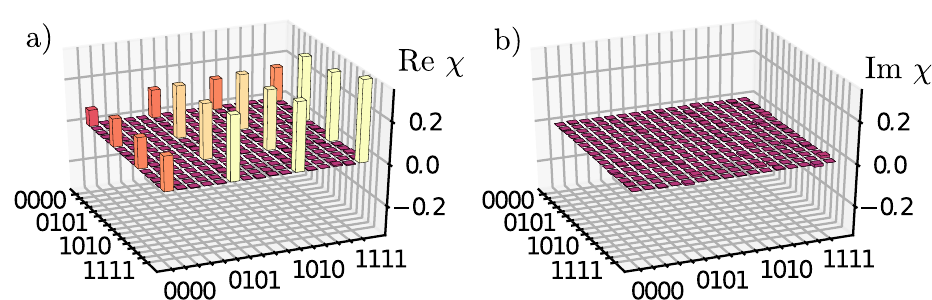}}
	\caption{Quantum process tomography of a quantum filter. Shown are the real (a) and imaginary (b) parts of the reconstructed Choi matrix $\hat{\chi}$ of experimentally implemented filter with nominal parameters $a=0.32$ and $b= 0.8$. 
		Residual phase shifts of the basis states $|jk\rangle$ were compensated before plotting the matrix $\hat{\chi}$.}
	\label{fig-filtertomo}
\end{figure}

\section{Experimental characterization of two-qubit quantum filters}

The quantum filter is a trace-decreasing completely positive map described by positive semidefinite 
Choi matrix $\hat{\chi}$ which can be obtained by applying the quantum filter to one part of a maximally entangled state in an extended Hilbert space of four qubits. In our experiments, 
we observe that besides amplitude modulation, the setup also imposes small residual phase shifts $\varphi_{jk}$ to the input states $|jk\rangle$.
Since these fixed phase shifts do not influence the ability of the filters to increase the mean energy or coherence, we compensate for these phase shifts in our tomographic analysis of the filter operation. 
An example of the tomographically reconstructed quantum filter is given in Fig~\ref{fig-filtertomo}. 
We observe the expected structure of a filter diagonal in the basis $|jk\rangle$ and the imaginary elements practically vanish.

\begin{table}[t!]
	\begin{tabular}{ll|lll}
		\hline
		$a$      & $b$      & $\mathcal{P}_M$        & $F_{M}$       & $\tilde{F}_{M}$\\
		\hline
		0.000~~  & 0.000~~  & 0.989(1)~~ & 0.9919(3)~~ & 0.9919(3) \\
		0.000  & 1.000  & 0.959(6) & 0.817(4) & 0.965(3) \\
		0.320  & 0.800  & 0.965(4) & 0.884(2) & 0.976(2) \\
		0.640  & 0.800  & 0.977(3) & 0.903(1) & 0.986(1) \\
		1.000  & 1.000  & 0.977(3) & 0.875(2) & 0.987(2)\\
		\hline
	\end{tabular}
	\caption{Characteristics of experimentally implemented quantum filters. The filter purity $\mathcal{P}$, fidelity $F_M$ and fidelity $\tilde{F}_M$ after compensation of phase shifts of basis states $|jk\rangle$ are shown for $5$ different nominal values of parameters $a$ and $b$. 
		Numbers in parenthesis represent one standard deviation. 
	}
	\label{table-filters}
\end{table}

A quantitative characterization of several implemented filters is provided in Table~\ref{table-filters}, where we display the filter purity $\mathcal{P}_M=\mathrm{Tr}[\hat{\chi}^2]/(\mathrm{Tr}[\hat{\chi}])^2$ and fidelity
$F_M=\mathrm{Tr}[\hat{\chi} \hat{\chi}_M]/(\mathrm{Tr}[\hat{\chi}]\mathrm{Tr}[\hat{\chi}_M])$. Here $\hat{\chi}_M$ denotes the Choi matrix of the perfect filter $\hat{M}$. This matrix has rank one and is proportional to a projector onto pure state. For perfect filters, 
we thus have $\mathcal{P}=F_M=1$. For reference, we display in Table~\ref{table-filters} also the fidelity $\tilde{F}_M$ 
calculated from the Choi matrix $\hat{\chi}$ with compensated phase shifts of basis states $|jk\rangle$. The main experimental factors that reduce the fidelity and purity of the linear optical filters include imperfect visibility 
of two-photon interference on the partially polarizing beam splitter PPBS, phase fluctuations and phase drift in the Mach-Zehnder interferometers, non-zero reflectance of PPBS for horizontally polarized states, and imperfections of the employed wave-plates.

\end{document}